# Mining User/Movie Preferred Features Based on Reviews for Video Recommendation System

**Xuan-Son Vu**[†]     **Seong-Bae Park**[†]

[†]School of Computer Science
Kyungpook National University, South Korea
{sonvx,sbpark}@sejong.knu.ac.kr

## Abstract

In this work, we present an approach for mining user preferences and recommendation based on reviews. There have been various studies worked on recommendation problem. However, most of the studies beyond one aspect user generated-content such as user ratings, user feedback and so on to state user preferences. There is a problem in one aspect mining is lacking for stating user preferences. As a demonstration, in collaborative filter recommendation, we try to figure out the preference trend of crowded users, then use that trend to predict current user preference. Therefore, there is a gap between real user preferences and the trend of the crowded people. Additionally, user preferences can be addressed from mining user reviews since user often comment about various aspects of products. To solve this problem, we mainly focus on mining product aspects and user aspects inside user reviews to directly state user preferences. We also take into account Social Network Analysis for cold-start item problem. With cold-start user problem, collaborative filter algorithm is employed in our work. The framework is general enough to be applied to different recommendation domains. Theoretically, our method would achieve a significant enhancement.

**Keywords:** Opinion Mining, Sentiment Aspect Analysis, System Recommendation

## 1 Introduction

User generated content in the form of comments has witnessed an explosive growth on the web. Websites like Amazon.com, IMDb.com allow users to comment on diverse contents like news articles. These "crowd-sourced" comments are highly engaging because they reflect the views and opinions of real users. Moreover, as the statistic from [1] 70% customers consult reviews or ratings before purchasing. Along with this, viewers also check movie reviews before making decision to buy movie tickets. There have been very few studies about movie recommendation that employ movie reviews. Therefore, in this work, we focus on mining user reviews to state user preferences for recommendation.

[1] http://www.businessweek.com

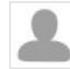

Figure 1: A user review on IMDb.com

However, the key challenges with user generated comments is that inside a review, people praise and criticize various aspects of the target such as the noise level of a computer or the taste of a dish. In the Figure 1, the reviewer evaluates aspects of a new blockbuster film called "Captain Philips" such as director, actors, movie scenes. Based on different aspect, the reviewer has different opinion. Therefore, user reviews are sufficient for stating user preference for recommendation. However, reviews are plain text without any structure, therefore how to mining aspects in reviews is a big challenge.

In this work, we propose a new approach for movie recommendation based on mining user reviews to state movie aspects and user preferences based on aspects. Generally, we apply LDA for finding hidden aspects for addressing user preference aspects and movie feature aspects. After user preferences based on aspects and movie aspects are addressed, KL divergence is used for measuring similarity between movie and user. Top K movies that close to user preferences are recommended to user.

Our main contributions in this work are including: (1) proposing a method to determine user opinion on aspects, (2) solving cold-start item recommendation by using Social Network Analysis to enrich item information, (3) applying collaborative filtering into cold-start user problem by finding the co-related opinion between massive users, and (4) publishing a first large review dataset with rating and comments of large anonymous reviewers on IMDb.

The rest of the paper is laid out as follows. Section 2 introduces some related works. Section 3, we address our approach to take advantage of topic modeling method to do recommendation. The Social Network Analysis along with Collaborative Filtering to overcome the cold-start item and cold-start user problem, respectively is deeply demonstrated in Section 4. Experiment is shown in Section 5. Section 6 gives conclusion and our future studies.

## 2 Related Work

In this section, we will show here the previous studies on mining user aspects, some studies by using movie reviews for recommendation and showing how are they different from our work.

There have been some studies about taking into account reviews for recommendation including [1; 2; 3]. In which, the most closed approach to ours is the work from Hariri et al. [1], that presented labeled-LDA to infer contexts based on review mining and combining them with user rating history for recommendation. However, their work requires supervised data that is difficult to achieve. Overall, there are three significant differences in our work from others including: (1) in our work, LDA is employed to state preferred features for both user and movie based on reviews, (2) we address and solve cold-start problems in our approach, and (3) we empirically apply our approach to the first large collected user reviews for recommendation.

## 3 Proposed Framework

In this section, we will briefly describe about our method. Due to page limitation of this short paper, we cannot show our approach in more detail.

### 3.1 Framework overview

The workflow of our system is deeply presented in figure 2 which consists of 8 major processing steps: 1) modeling overall Model for all reviews in corpus; 2) getting all user-related reviews in corpus; 3) inference user feature preferences by using global model; 4) using collaborative filter between user to find cross related-preferences to solve cold-start user problem; 5) getting Tweets information for cold-start movie; 6) inference movies' feature preference by using global model achieved in step 2; 7) using similarity measurement (e.g cosine similarity) to find the similarity between user feature preferences and movie features; and 8) top K movies is generated to recommend to the user.

### 3.2 Stating User Feature Preferences

For users who has sufficient information (enough reviews), we will construct their feature preferences from their reviews to the movies. Basically, when a user writes a comment, s/he will comment about their opinions on item features that called feature-opinion pairs in [2]. However, in [2], Feng et al. discover features of item based on Part-of-Speech that will decrease the accuracy of detected-features since a lot of words with same tag but they are not feature of item. Therefore, in our work, LDA is employed to detect those features. Obviously some noised features are also detected by LDA but their proportion will be lower than the true features since preprocessing process. For the cold-start user, we will deeply address in the subsection 4.1.

### 3.3 Stating Movie Feature Preferences

All reviews of a movie are consider as a unseen document to the global topic model. By referring from global topic model, movie preferred features are achieved. For the movie that has limited reviews, we solve it by taking into account social network information as addressed in the subsection 4.2.

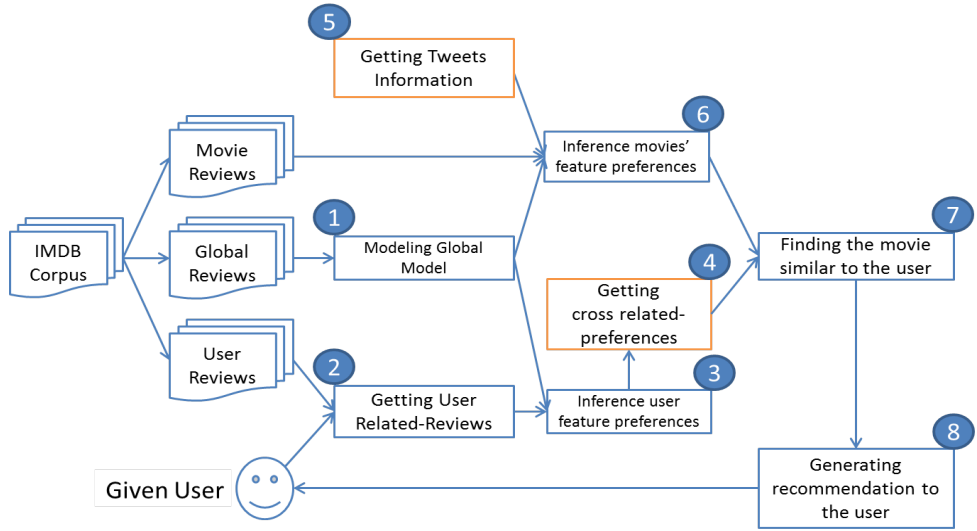

Figure 2: **System workflow**

## 4 Cold-start problem

### 4.1 Cold-start User Recommendation

Since our approach is mainly focusing on Recommendation for IMDb users by taking into account their reviews data. Therefore our cold-start user in this case is the user that has insufficient review (e.g: User has one short review). Basically, we adopt collaborative filtering method of user-item to user-feature for finding cross related-features between user. So that the cold-start user will have sufficient information for recommendation. In more detail, we will describe on the full paper.

### 4.2 Cold-start Movie Recommendation

In our work, the problem of cold-start movie is for incoming movies that will be released in next few weeks/months. Since the review information is insufficient for applying recommendation. We found that with the incoming movies, the producers have been posted many information about the movies on Twitter. Additionally there are many Twitter users follow the movies and give some Tweets on the movies. The table 1 shows the domination of Tweets information over IMDb reviews within one month from released date of the movie "American Hustle". In [4], Lin et al. also show that despite the cold-start, there is still information out there about the item, particularly on social networking services like Twitter. Therefore we collect user tweets about cold-movies on Twitter for suffi-cient movie reviews. This step will be described in more detail in our full paper.

| IMDb | Tweet |
|---|---|
| 16 | 331 |
| 352 | 591 |

Table 1: Statistic of the collected Reviews/Tweets within 1 month.

| Criteria | Number |
|---|---|
| Number of Movie | 14,127 |
| Number of User | 277,490 |
| Number of Review | 677,675 |
| Number of Rating | 542,348 |

Table 2: Statistic of the collected dataset.

## 5 Experiment

### 5.1 Dataset

In [5], Andrew et al. published the first large movie reviews dataset which contains 50,000 reviews with corresponded rating information. However, there is no information about holder (a person that wrote the review) for the related reviews. Therefore, by using the IMDb URLs that attached along with the corpus, we re-collected data information that including reviews, rating, and the holder identity. The statistic of the collected dataset is shown in figure 2.

### 5.2 Exploiting Movie Preferred Features

The figure 3 shows three first preferred features of movie reviews in the dataset. In here, LDA with Gibb sampling of Phan et al. [6] is employed to exploit movie preferred features with 50 topics in 1000 iterations.

| actor | action | animation |
|---|---|---|
| william 0.0140025 | comic 0.01513 | anim 0.06865 |
| perform 0.012570 | superman 0.01430 | cartoon 0.02099 |
| john 0.01217059 | movi 0.013123 | voic 0.01523 |
| best 0.0111586 | film 0.008805 | charact 0.0141 |
| oscar 0.0089178 | charact 0.00849 | stori 0.00994 |
| great 0.0082286 | batman 0.00816 | watch 0.00618 |
| actor 0.007091 | like 0.00769 | human 0.00478 |
| hank 0.006953 | good 0.007589 | children 0.004651 |
| hoffman 0.006236 | action 0.007455 | pixar 0.004011 |

Figure 3: Top word distribution in top three movie preferred features detected by LDA with Gibb sampling.

### 5.3 Evaluation Methodology

The dataset as described in 5.1 is used to perform the experiment. Given that a user has a target movies to watch (which is her/his target choices), the experimental goal is to evaluate whether the target choices can be located in the recommendation list when being presented to her/him. For this goal, we concretely adopted the leave-one out evaluation scheme [3]. That is, during each round, we excluded one reviewer from the dataset and performed testing on it. As a matter of fact, the excluded reviewer must satisfy two "new user" selection criteria, so that the product purchased by the reviewer can be taken as the new user's target choice when measuring the algorithm's recommendation accuracy: 1) s/he has top number of reviews over the other users, and 2) her/his rating on the watched movies is positive marks (i.e., from 7 to 10), indicating that s/he likes the movies. In our dataset, we designed to select 1000 reviewers with the above criteria. Therefore, at a time, one of them will be randomly chosen to behave as a new user. We further randomly select subsets of the reviewer's full feature preferences (i.e., 40%, 60%, 80% and 100%) to represent the new user's various preference completeness levels.

## 6 Conclusion

In this paper, we have presented a thorough investigation of using user reviews for recommendation task. Our key contributions of the work is four-fold including (1) proposing a method to determine user opinion on aspects, (2) solving cold-start item recommendation by using Social Network Analysis to enrich item information, (3) applying collaborative filtering into cold-start user problem by finding the co-related opinion between massive users, and (4) publishing a first large review dataset with rating and comments of large anonymous reviewers on IMDb.

About future study, firstly we will perform the evaluation method to proof our approach. Secondly, we manage to propose a generative model for exploiting user preferred features and movie preferred features at once. So that we can reduce the complexity of referring steps and exploit features in advance.

web with hidden topics from large-scale data collections. pages 91–100, 2008.